\documentclass{emulateapj}

\usepackage{amssymb}
\usepackage{amsmath}
\usepackage{epstopdf}
\usepackage{natbib}
\usepackage[colorlinks, citecolor=blue]{hyperref}
\usepackage{hyperref}
\usepackage[all]{hypcap}

\begin{document}

\title{A Multiple Ejecta-Circumstellar Meduim Interaction Model and Its
Implications for the Superluminous Supernova \lowercase
{i}PTF15\lowercase{esb}}
\author{Liang-Duan Liu\altaffilmark{1,2}, Ling-Jun Wang\altaffilmark{3},
Shan-Qin Wang\altaffilmark{1,2,4}, Zi-Gao Dai\altaffilmark{1,2}}

\begin{abstract}
Recently, a hydrogen-poor superluminous supernova (SLSN) iPTF15esb at
redshift $z=0.224$ was reported, whose light curve (LC) and spectrum show
several unusual characteristics. Its late-time spectrum shows a strong,
broad H$\alpha $ emission line and the bolometric LC exhibits two peaks and
a post-peak plateau. Here we propose an ejecta-circumstellar medium (CSM)
interaction model involving multiple shells and winds to explain this
double-peak SN. We find that the theoretical LC reproduced by this model can
well match the observations of iPTF15esb. Based on this result, we infer
that the progenitor has undergone at least three violent mass-loss processes
before the SN explosion. Furthermore, we find that the masses of the CSM
wind and shells from the outermost shell (the first eruption) to the
innermost wind (the final eruption) decrease sequentially but their
densities increase. The variation trend of the inferred densities of the
shells and wind is consistent with the stellar structure before an SN
explosion. Further investigations for similar SLSNe would provide a probe
for the mass-loss history of their progenitors.
\end{abstract}


\affil{\altaffilmark{1}School of Astronomy and Space Science,
Nanjing University, Nanjing 210093, China; dzg@nju.edu.cn}
\affil{\altaffilmark{2}Key Laboratory of Modern Astronomy and
Astrophysics (Nanjing University), Ministry of Education, China}
\affil{\altaffilmark{3}Key Laboratory of Space Astronomy and
Technology, National Astronomical Observatories, Chinese Academy of
Sciences, Beijing 100012, China}
\affil{\altaffilmark{4}Department
of Astronomy, University of California, Berkeley, CA 94720-3411,
USA}

\section{INTRODUCTION}

\label{sec:Intro}

In the past decade, fast-developing non-targeted supernova (SN) survey
programs have discovered a new class of unusual SNe whose peak absolute
magnitudes $M_{\mathrm{peak}}$ at all bands are brighter than $-21$ mag.
These very luminous SNe are called \textquotedblleft superluminous
supernovae (SLSNe)" \citep{Qiu2011,Gal2012}.

It appears that SLSNe can be simply divided into two categories, SLSNe I and
II. SLSNe I have spectra around the peaks that are lack of hydrogen
absorption lines and their light curves (LCs) might be explained by the pair
instability SNe \citep[PISNe;][]{Rak1967,Heg2002,Heg2003} model (i.e., $%
^{56} $Ni model), the magnetar model %
\citep{Kas2010,Woo2010,Ins2013,Wang2015a,Wang2015b,Dai2016,Wang2016a,Liu2017,Yu2017}
or the ejecta-circumstellar (CSM) interaction model %
\citep{Che2011,Gin2012,Cha2012,Cha2013,Nic2014,Chen2015}.

On the other hand, the spectra around the peaks of SLSNe II show strong
hydrogen emission features and almost all of them show narrow and
intermediate width Balmer emission lines, similar to normal SNe IIn.
Previous studies \citep{Sim2007,Mor2011,Cha2012,Cha2013,Mor2013} suggested
that the LCs of SLSNe IIn might be powered by the interactions between the
SN ejecta and the dense, hydrogen-rich, and optically thick CSM.

However, some SLSNe I (e.g., iPTF13ehe, iPTF15esb and iPTF16bad) whose
late-time spectra exhibit H$\alpha $ emission lines \citep{Yan2015,Yan2017}
complicated the classification scheme. \citet{Yan2015} estimated that $15\%$
of SLSNe I might have these spectral features. Among these SLSNe I that have
late-time H$\alpha $ emission lines, iPTF15esb, exploded at redshift $%
z=0.224 $, is the most striking one. Its late-time spectra show strong,
broad H$\alpha $ emission lines, indicative of the interaction between the
SN ejecta and the hydrogen-rich CSM shell surrounding the SN progenitor.
Moreover, its LC has two peaks whose luminosities are approximately equal to
each other ($L_{\text{peaks}}\approx 4\times 10^{43}$ erg s$^{-1}$) and a
plateau lasting about 40 days. Its late-time LC decays as $L_{\text{bol}%
}\propto t^{-2.5}$.\footnote{%
As pointed out by \citet{Yan2017}, undulation features that clearly deviate
from the smooth rise and fall are also seen in the LCs of other SLSNe I such
as SN2015bn \citep{Nic2016} and iPTF13dcc \citep{Vre2017}.}

The decline rate of the late-time LC powered by $^{56}$Ni cascade decay with
full trapping of $\gamma $-rays is 0.0098 mag per day and the late-time LCs
powered by a magnetar (with full trapping of $\gamma $-rays) can be
described by $L_{\text{inp,mag}}\propto t^{-2}$. The magnetar model together
with $^{56}$Ni cascade decay with leakage of $\gamma $-rays %
\citep{Clo1997,Wang2015a,Chen2015} is able to explain the late-time LC of
iPTF15esb. However, neither $^{56}$Ni cascade decay model nor magnetar model
can produce the LC showing two bright peaks and a plateau.

It seems that an energy-source model involving multiple energy injections is
needed to account for the exotic LC of iPTF15esb. \cite{Wang2016b} propose a
triple-energy-source model (i.e., $^{56}$Ni plus magnetar plus interaction)
for iPTF13ehe. However, this model involve only one collision between the
ejecta and the CSM shell and also cannot produce an LC having two peaks and
a long-lasting plateau.

Therefore, the LC of iPTF15esb cannot be explained by the models mentioned
above. As noted by \cite{Yan2017}, however, the spectrum and the LC might
seem to favor the interactions between the SN ejecta and multiple CSM shells
or CSM clumps at different radii. Here we propose an ejecta-CSM interaction
scenario involving interactions between the SN ejecta and multiple shells
and stellar winds and adopt this scenario to model the whole LC of
iPTF15esb. The paper is structured as follows. In Section \ref{sec:mod}, we
give a detailed description of the model, and apply it to fit the LC and
temperature evolution of iPTF15esb in Section \ref{sec:res}. Finally, we
discuss our results and conclude in Section \ref{sec:dis}.

\section{Multiple Ejecta-CSM Interaction Model}

\label{sec:mod}

In this section, we generalize the normal ejecta-CSM interaction model to a
model involving multiple CSM shells and winds. The basic physical picture of
this model will be described below.

The interaction of the ejecta with the pre-existing CSM results in the
formation of two shock waves: a forward shock (FS) propagating through the
CSM and a reverse shock (RS) sweeping up the SN ejecta %
\citep{Che1982,Che1994}. The interaction provides a strong energy source by
the conversion of kinetic energy into radiation.

Based on the numerical simulations for SN explosions, a double power-law
distribution for the density of the SN ejecta can be adopted \citep{Mat1999}%
. The density profile of the outer ejecta is
\begin{equation}
\rho _{\text{NS,out}}=g_{n}t^{n-3}r^{-n},
\end{equation}%
where $n$ is the slope of the outer ejecta, depending on the SNe progenitor
stars, and $g_{n}$ is the density profile scaling parameter, which is given
by \citep{Che1994,Cha2012}
\begin{equation}
g_{n}=\frac{1}{4\pi \left( n-\delta \right) }\frac{\left[ 2\left( 5-\delta
\right) \left( n-5\right) E_{\text{SN}}\right] ^{\left( n-3\right) /2}}{%
\left[ \left( 3-\delta \right) \left( n-3\right) M_{\text{ej}}\right]
^{\left( n-5\right) /2}},
\end{equation}
where $\delta $ is the inner density profile slope. Here $E_{\text{SN}}$ is
the total SN energy, and $M_{\text{ej}}$ is the total mass of the SN ejecta.
The relation between $E_{\text{SN}}$ and $M_{\text{ej}}$ could be written as %
\citep{Cha2012}
\begin{equation}
E_{\text{SN}}=\frac{3\left( n-3\right) }{2\left( 5-\delta \right) \left(
n-5\right) }M_{\text{ej}}\left( x_{0}v_{\text{SN}}\right) ^{2},
\end{equation}
where $x_0$ denotes the dimensionless radius of break in the supernova
ejecta density profile from the inner component to the outer component.

Before the SN explosion, the mass loss of massive stars could erupt several
gas shells. We assume that the circumstellar density follows
\begin{equation}
\rho _{\text{CSM,i}}=q_{\text{i}}r^{-s_{\text{i}}},
\end{equation}%
where $q$ is a scaling constant, and $s$ is the power-law index for CSM
density profile and therefore $s=2$ indicates stellar winds while $s=0$
indicates uniform density shells. The subscript \textquotedblleft i" denotes
the $i$th collision between the ejecta and the CSM shell. For a steady wind (%
$s=2$) with a constant pre-explosion mass loss rate $\dot{M}$ and wind
velocity $v_{\text{w}}$, we have $q=\dot{M}/\left( 4\pi v_{\text{w}}\right) $%
.

The shocked CSM and shocked ejecta are separated by a contact discontinuity.
The radius of contact discontinuity $R_{\text{cd}}$ can be described by a
self-similar solution \citep{Che1982}
\begin{equation}
R_{\text{cd,i}}=\left( \frac{A_{\text{i}}g_{n}}{q_{\text{i}}}\right) ^{\frac{%
1}{n-s_{\text{i}}}}t^{\frac{\left( n-3\right) }{\left( n-s_{\text{i}}\right)
}},
\end{equation}%
where $A$ is a constant. The radii of the FS and the RS are given by
\begin{equation}
R_{\text{FS,i}}\left( t\right) =R_{\text{in,i}}+\beta _{\text{FS,i}}R_{\text{%
cd,i}}
\end{equation}%
and
\begin{equation}
R_{\text{RS},\text{i}}\left( t\right) =R_{\text{in,i}}+\beta _{\text{RS,i}%
}R_{\text{cd,i}},
\end{equation}%
where $R_{\text{in}}$ is the interaction radius (equal to the inner radius
of CSM density profile), $\beta _{\text{FS}}$ and $\beta _{\text{RS}}$ are
constants representing the ratio of the shock radii to the
contact-discontinuity radius $R_{\text{cd}}$. The values of $\beta _{\text{FS%
}}$ and $\beta _{\text{RS}}$ are determined by the values of $n$ and $s$.
They are given in Table 1 of \cite{Che1982}. For $n=7$ and $s=2$, we can
obtain $\beta _{\text{FS}}=1.299$, $\beta _{\text{RS}}=0.970$, and $A=0.27$;
for $n=7$ and $s=0$, we have $\beta _{\text{FS}}=1.181$, $\beta _{\text{RS}%
}=0.935$, and $A=1.2$.

The interaction radii of the second and third collisions are given by
\begin{equation}
R_{\text{in,2}}=R_{\text{in,1}}+\left( t_{\text{shift,2}}-t_{\text{shift,1}%
}\right) \left( \frac{2\left( 5-\delta \right) \left( n-5\right) E_{\text{k,2%
}}}{3x_{0}^{2}\left( n-3\right) M_{\text{ej,2}}}\right) ^{1/2},
\label{eq: Rin1}
\end{equation}%
and
\begin{equation}
R_{\text{in,3}}=R_{\text{in,2}}+\left( t_{\text{shift,3}}-t_{\text{shift,2}%
}\right) \left( \frac{2\left( 5-\delta \right) \left( n-5\right) E_{\text{k,3%
}}}{3x_{0}^{2}\left( n-3\right) M_{\text{ej,3}}}\right) ^{1/2},
\label{eq: Rin2}
\end{equation}%
where $t_{\text{shift}}$ is the trigger time of interaction relative to time
zero point. Here, we set the first peak of the LC as time zero point. The
kinetic energies of the second and third interactions are
\begin{equation}
E_{\text{k,2}}=E_{\text{k,1}}-E_{\text{rad,1}},\text{ \ \ \ \ \ }E_{\text{k,3%
}}=E_{\text{k,2}}-E_{\text{rad,2}},
\end{equation}%
where $E_{\text{rad}}$ is the energy loss due to radiation.

The interaction between the ejecta and the CSM would convert the kinetic
energy to radiation. Using energy conservation, \cite{Cha2012} found that
the luminosity input functions from the FS and RS are
\begin{eqnarray}
L_{\text{FS,i}}\left( t\right)  &=&\frac{2\pi }{\left(
n-s_{\text{i}}\right)
^{3}}g_{n}^{\frac{5-s_{\text{i}}}{n-s_{\text{i}}}}q_{\text{i}}^{\frac{n-5}{%
n-s_{\text{i}}}}\left( n-3\right) ^{2}\left( n-5\right) \beta _{\text{FS,i}%
}^{5-s_{\text{i}}}  \notag \\
&&\times A^{\frac{n-5}{n-s_{\text{i}}}}\left(
t+t_{\text{int,i}}\right) ^{\alpha _{\text{i}}}\theta \left(
t_{\text{FS},\text{BO,i}}-t\right)
\end{eqnarray}
and
\begin{eqnarray}
L_{\text{RS,i}}\left( t\right)  &=&2\pi \left( \frac{A_{\text{i}}g_{n}}{q_{%
\text{i}}}\right) ^{\frac{5-n}{n-s_{\text{i}}}}\beta _{\text{RS,i}%
}^{5-n}g_{n}\left( \frac{3-s_{\text{i}}}{n-s_{\text{i}}}\right) ^{3}
\notag
\\
&&\times \left( t+t_{\text{int,i}}\right) ^{\alpha
_{\text{i}}}\theta \left( t_{\text{RS},\ast \text{,i}}-t\right) ,
\end{eqnarray}

where $\theta \left( t_{\text{RS},\ast }-t\right) $ and $\theta \left( t_{%
\text{FS},\text{BO}}-t\right) $ represent the Heaviside step function that
controls the end times of FS and RS, respectively. $t_{\text{int}}\approx R_{%
\text{in}}/v_{\text{SN}}$ is the time when the ejecta-CSM interaction
begins. The temporal index is $\alpha =\left( 2n+6s-ns-15\right) /\left(
n-s\right) $. Here we fix $n=7$. Consequently, we have $\alpha =-0.143$ for
the shells ($s=0$), and $\alpha =-0.6$ for the steady winds ($s=2$).

Once the RS sweeps up all available ejecta, the RS termination timescale $t_{%
\text{RS},\ast }$ can be obtained \citep{Cha2012,Cha2013}
\begin{equation}
t_{\text{RS,}\ast ,\text{i}}=\left[ \frac{v_{\text{SN,i}}}{\beta _{\text{RS,i%
}}\left( A_{\text{i}}g_{n}/q_{\text{i}}\right) ^{\frac{1}{n-s_{\text{i}}}}}%
\left( 1-\frac{\left( 3-n\right) M_{\text{ej,i}}}{4\pi v_{\text{SN,i}%
}^{3-n}g_{n}}\right) ^{\frac{1}{3-n}}\right] ^{\frac{n-s_{\text{i}}}{s_{%
\text{i}}-3}}.
\end{equation}%
The masses of ejecta of the second and third interactions are
\begin{equation}
M_{\text{ej,2}}=M_{\text{ej,1}}+M_{\text{CSM,1}},\text{ \ \ }M_{\text{ej,3}%
}=M_{\text{ej,2}}+M_{\text{CSM,2}}.  \label{eq: Mej}
\end{equation}%
Under the same assumption, the FS terminates when the optically thick part
of the CSM is swept up. The termination timescale of the FS, being
approximately equal to the time of FS breakout $t_{\text{FS},\text{BO}}$, is
given by \citep{Cha2012,Cha2013}
\begin{eqnarray}
t_{\text{FS,BO,i}} &=&\left\{ \frac{\left( 3-s_{\text{i}}\right) q_{i}^{%
\frac{3-n}{n-s_{\text{i}}}}\left[ A_{\text{i}}g_{n}\right] ^{\frac{s_{\text{i%
}}-3}{n-s_{\text{i}}}}}{4\pi \beta
_{\text{FS,i}}^{3-s_{\text{i}}}}\right\}
^{\frac{n-s_{\text{i}}}{\left( n-3\right) \left(
3-s_{\text{i}}\right) }}
\notag \\
&&\times M_{\text{CSM,th,i}}^{\frac{n-s_{\text{i}}}{\left(
n-3\right) \left( 3-s_{\text{i}}\right) }}
\end{eqnarray}
where $M_{\text{CSM,th}}$ is the mass of optically thick CSM
\begin{equation}
M_{\text{CSM,th,i}}=\int_{R_{\text{in,i}}}^{R_{\text{ph,i}}}4\pi r^{2}\rho _{%
\text{CSM,i}}dr.
\end{equation}%
Here $R_{\text{ph,i}}$ denotes the photospheric radius of the $i$th CSM
shell, located at the optical depth $\tau =2/3$ under Eddington's
approximation. $R_{\text{ph}}$ is given by
\begin{equation}
\tau =\int_{R_{\text{ph,i}}}^{R_{\text{out,i}}}\kappa _{\text{i}}\rho _{%
\text{CSM,i}}dr=\frac{2}{3},
\end{equation}%
where $\kappa $ is the optical opacity of the CSM and $R_{\text{out}}$ is
the radius of outer boundary of CSM. $R_{\text{out}}$ can be determined by
the following expression:
\begin{equation}
M_{\text{CSM,i}}=\int_{R_{\text{in,i}}}^{R_{\text{out,i}}}4\pi r^{2}\rho _{%
\text{CSM,i}}dr.
\end{equation}

Both the FS and the RS heat the interacting material. The total luminosity
input from the FS and RS can be written as
\begin{equation}
L_{\text{inp,CSM,i}}\left( t\right) =\epsilon _{\text{i}}\left[ L_{\text{FS,i%
}}\left( t\right) +L_{\text{RS,i}}\left( t\right) \right] ,
\end{equation}%
where $\epsilon $ is the conversion efficiency from the kinetic energy. \cite%
{Cha2012} assumed that $\epsilon =$ 100$\%$, which is unrealistic in the
actual situation, especially in the $M_{\text{CSM}}\ll M_{\text{ej}}$ case.
Due to the poor knowledge of the process of converting the kinetic energy to
radiation, for simplicity, we set $\epsilon $ as a free parameter.

Because the expansion velocity of the CSM is much lower than the typical
velocity of SN ejecta, \cite{Cha2012} assumed a fixed photosphere inside the
CSM. Under this assumption, the output bolometric LC can be written as
\begin{equation}
L_{\text{i}}\left( t\right) =\frac{1}{t_{\text{diff,i}}}\exp \left[ -\frac{t%
}{t_{\text{diff,i}}}\right] \int_{0}^{t}\exp \left[ \frac{t^{\prime }}{t_{%
\text{diff,i}}}\right] L_{\text{inp,CSM,i}}\left( t^{\prime }\right)
dt^{\prime },
\end{equation}%
where $t_{\text{diff}}$ is the diffusion timescale in the optically thick
CSM. The diffusion timescales of three interactions can be written as
\begin{eqnarray}
t_{\text{diff,1}} &=&\frac{\kappa _{1}M_{\text{CSM,th,1}}+\kappa _{2}M_{%
\text{CSM,th,2}}+\kappa _{3}M_{\text{CSM,th,3}}}{\beta cR_{\text{ph}}},
\notag \\
t_{\text{diff,2}} &=&\frac{\kappa _{2}M_{\text{CSM,th,2}}+\kappa _{3}M_{%
\text{CSM,th,3}}}{\beta cR_{\text{ph}}}, \\
t_{\text{diff,3}} &=&\frac{\kappa _{3}M_{\text{CSM,th,3}}}{\beta cR_{\text{ph%
}}},  \notag
\end{eqnarray}%
where $\beta =4\pi ^{3}/9\simeq 13.8$ is a constant \citep{Arn1982}, and $c$
is the speed of light. It is worth noting that there is only one
photospheric radius in our model, i.e., $R_{\text{ph}}=R_{\text{ph,3}}$.

Based on this multiple ejecta-CSM interaction model, we can obtain the
theoretical bolometric LC
\begin{equation}
L_{\text{tot}}\left( t\right) =L_{1}\left( t+t_{\text{shift,1}}\right)
+L_{2}\left( t+t_{\text{shift,2}}\right) +L_{3}\left( t+t_{\text{shift,3}%
}\right) .
\end{equation}

We assume that the bolometric luminosity come from the blackbody emission
from the photosphere whose radius is $R_{\text{ph}}$, and therefore the
temperature in our model can be estimated by
\begin{equation}
T=\left( \frac{L_{\text{tot}}}{4\pi R_{\text{ph}}^{2}\sigma }\right) ^{1/4},
\end{equation}%
where $\sigma $ is the Stefan-Boltzmann constant. By assuming a stationary
photosphere, we have $T\propto L_{\text{tot}}^{1/4}$.

\begin{table*}[tbph]
\caption{Fitting Parameters for {\upshape iPTF15esb}}
\label{tbl:fitting par}
\begin{center}
\begin{tabular}{ccccccccc}
\hline\hline
$i$th interaction & $s$ & $\kappa $ & $M_{\text{ej}}$ & $M_{\text{CSM}}$ & $%
\rho _{\text{CSM,in}}$ \tablenotemark{b} & $\epsilon $ \tablenotemark{c} & $%
t_{\text{shift}}$ & $R_{\text{in}}$ \\
&  & $\left( \text{cm}^{2}\text{ g}^{-1}\right) $ & $\left( M_{\odot
}\right) $ & $\left( M_{\odot }\right) $ & $\left( 10^{-13}\text{ g cm}%
^{-3}\right) $ &  & $\left( \text{days}\right) $ & $\left( 10^{15}\text{cm}%
\right) $ \\ \hline
1 & $2$ & 0.2 & 6.1 & 0.42 & 22 & 0.29 & $-8.5$ & 0.2 \\
2 & 0 & 0.2 & 6.32 \tablenotemark{a} & 1.32 & 4.5 & 0.13 & 5.2 & 2.1 %
\tablenotemark{d} \\
3 & 0 & 0.33 & 7.64 \tablenotemark{a} & 1.81 & 0.21 & 0.11 & 24.2 & 4.8 %
\tablenotemark{d} \\ \hline
\end{tabular}%
\end{center}
\par
a. $M_{\text{ej,2}}$ and $M_{\text{ej,3}}$ are not fitting
parameters, but calculated by Equation (\ref{eq: Mej}).
\par
b. $\rho_{\text{CSM,in}}$ the density of the CSM at the radius $R=R_{\text{in%
}}$.
\par
c. $\epsilon $ is the conversion efficiency from the kinetic energy
to radiation.
\par
d. $R_{\text{in,2}}$ and $R_{\text{in,3}}$ are not fitting
parameters, but calculated by Equation (\ref{eq: Rin1}) and
(\ref{eq: Rin2}).
\end{table*}

\section{Implications for iPTF15esb}

\label{sec:res}

In this section, we use the model described above to fit the bolometric LC
and temperature evolution of iPTF15esb. It is reasonable to suggest at least
three collisions between the SN ejecta and CSM shells since the LC of
iPTF15esb shows two prominent peaks and a plateau. In order to reduce the
number of free parameters in our calculations, we fix several parameters.
\begin{figure}[tbph]
\begin{center}
\includegraphics[width=0.53\textwidth,angle=0]{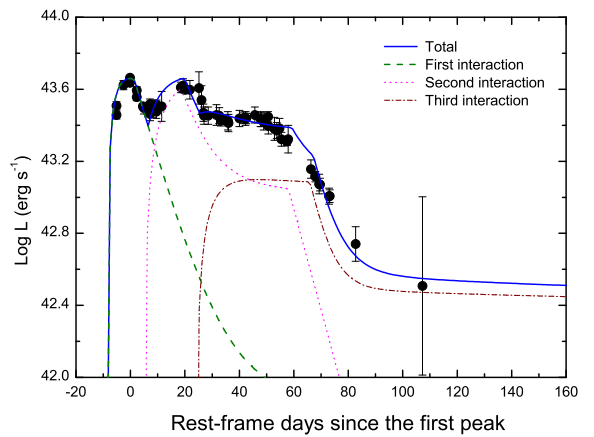}
\end{center}
\caption{The fit to the bolometric LC of iPTF15esb using the
multiple ejecta-CSM interaction model. Data are obtained from
\protect\cite{Yan2017}.
The fitting parameters are shown in the text and Table \protect\ref%
{tbl:fitting par}.} \label{fig:LC}
\end{figure}

In our model, we assume that the interaction between the SN ejecta and the
stellar wind (i.e., $s_{1}=2$) power the first peak of the LC of iPTF15esb.
The second peak and the plateau are powered by the interactions between the
SN ejecta and CSM shells at different radii, in which $s_{2}=s_{3}=0$ is
adopted.

By analyzing Fe II 5169 \AA\ line, \cite{Yan2017} found that the
photospheric velocity around the first peak of iPTF15esb is $v_{\text{ph}%
}\approx 17800$ km s$^{-1}$. Here we use $v_{\text{ph}}$ as the
characteristic velocity $v_{\text{SN}}$ of the ejecta. In addition, we adopt
the power-law index of the inner density profile $n=7$ as an approximation
for Type I SNe \citep{Che1982}, the inner density slope $\delta =0$, and the
dimensionless radius $x_{0}=0.3$.

The opacities of the CSM shells and winds $\kappa $ are related to their
composition and temperatures. For hydrogen-poor matter, and when the
dominant source of opacity is electron scattering, $\kappa =0.06-0.2$ cm$^{2}
$ g$^{-1}$ (see the references listed in \citealt{Wang2015b}). For
hydrogen-rich matter, $\kappa =0.33$\,cm$^{2}$\,g$^{-1}$, which is the Thomson
electron scattering opacity for fully ionized material with solar
metallicity \citep{Mor2011,Cha2012}. Since no hydrogen emission lines in the
early-time spectra of iPTF15esb have been detected, we have no idea about
the composition of the first and second CSM shells (from the spectra we may
expect them to be hydrogen-poor) and suggest that $\kappa _{1}=\kappa
_{2}=0.06-0.2$\,cm$^{2}$\,g$^{-1}$ or 0.33\,cm$^{2}$\,g$^{-1}$. There is a
strong, broad H$\alpha $ emission at $\sim 70$ days from the first light
peak. We interpret the H$\alpha $ emission as the result of ejecta
interaction with hydrogen-rich shell. Therefore, we adopt the opacity of
third CSM shell $\kappa _{3}=0.33$\,cm$^{2}$\,g$^{-1}$.

Thus, there are six free parameters in our model: the mass of the SN ejecta $%
M_{\text{ej}}$, the total mass of CSM $M_{\text{CSM}}$, the density of the
CSM at the interaction radius $\rho _{\text{CSM,in}}$, the interaction
radius (the inner radius of CSM) $R_{\text{in}}$, the conversion efficiency
from the kinetic energy to radiation $\epsilon $, the time of the collision
between the SN ejecta and the CSM shells $t_{\text{shift}}$. Most of the
model parameters for the second and third interactions are fixed by the
fitting parameters of the first interaction.

\begin{figure}[tbph]
\begin{center}
\includegraphics[width=0.53\textwidth,angle=0]{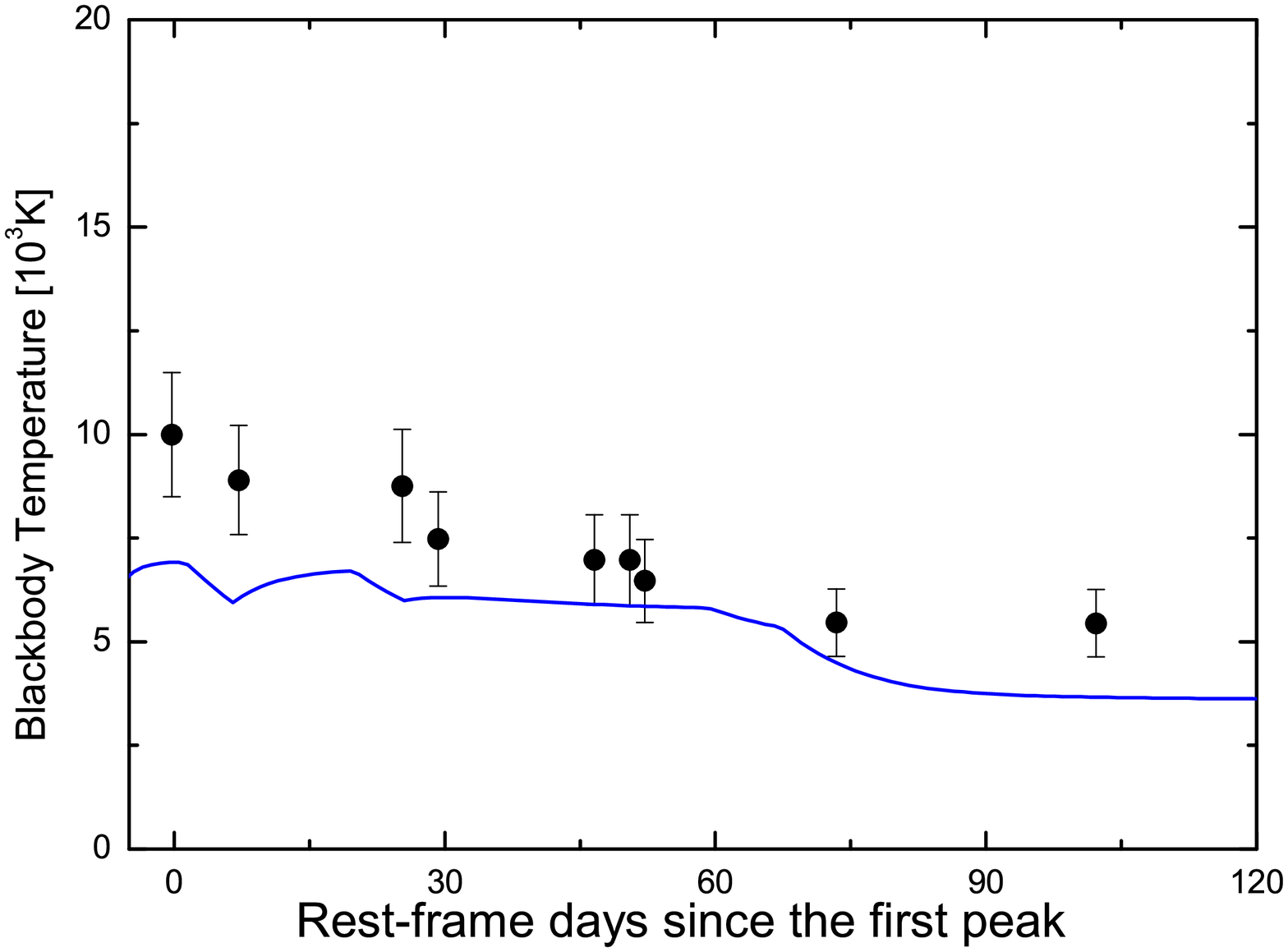}
\end{center}
\caption{The fit to the blackbody temperature evolution of iPTF15esb
using the multiple ejecta-CSM interaction model. Data are obtained
from \protect\cite{Yan2017}. The fitting parameters are shown in the
text and Table \protect\ref{tbl:fitting par}.} \label{fig:Tem}
\end{figure}

The theoretical LC and temperature evolution are shown in Figures \ref%
{fig:LC} and \ref{fig:Tem}, respectively. The corresponding parameters are
listed in Table \ref{tbl:fitting par}. We find that the multiple interaction
model can explain the special LC of iPTF15esb well and the parameters are
reasonable. In our model the blackbody radiation emanates from a fixed
photosphere. The assumption of several stationary CSM shells results in a
difference between the theoretical velocity evolution and observational data
at early times, as shown in Figure \ref{fig:Tem}, indicating that this
assumption ought to be modified in the case of a wind.

The physical parameters of the CSM shells and the wind are listed in Table %
\ref{tbl:der}. The mass of the optically thick part of the CSM shell $M_{%
\text{CSM,th}}$, which is close to their total mass, can be determined. The
termination timescales of the FS and the RS can also be determined. The
optical depth of CSM $\tau _{\text{CSM}}>1$, indicating that these shells
are opaque.

Provided that the velocity of the stellar wind $v_{\text{w}}$ and the
expansion velocities of the shells $v_{\text{shell}}$ are 1000 km s$^{-1}$
and 100 km s$^{-1}$, respectively, we can obtain the time when the
progenitor expelled the CSM shells before explosion by using $t_{\text{erupt}%
}\approx R_{\text{in}}/v_{\text{shell}}$. We infer that the progenitor has
undergone at least three violent mass-loss processes at 0.03, 6.9, and 15.5
years before the supernova explosion, respectively.

\begin{table*}[tbph]
\caption{Derived Physical Parameters} \label{tbl:der}
\begin{center}
\begin{tabular}{cccccccc}
\hline\hline $i$th interaction &
$M_{\text{CSM,th}}$\tablenotemark{a} & $t_{\text{FS,BO}}$
& $t_{\text{RS,*}}$ & $t_{\text{diff}}$ & $\tau _{\text{CSM}}$ %
\tablenotemark{b} & $R_{\text{out}}$ & $t_{\text{erupt}}$
\tablenotemark{c}
\\
& $\left( M_{\odot }\right) $ & $\left( \text{days}\right) $ & $\left( \text{%
days}\right) $ & $\left( \text{days}\right) $ &  & $\left( 10^{15}\text{ cm}%
\right) $ & $\left( \text{yr}\right) $ \\ \hline
1 & 0.40 & 7.1 & 278.6 & 8.5 & 69.6 & 0.96 & 0.06 \\
2 & 1.21 & 13.3 & 53.7 & 7.7 & 8.5 & 2.3 & 6.9 \\
3 & 1.45 & 41.7 & 165.5 & 5.2 & 4.1 & 5.3 & 15.5 \\ \hline
\end{tabular}%
\end{center}
\par
a. $M_{\text{CSM,th}}$ is the mass of optically thick CSM.
\par
b. $\tau _{\text{CSM}}$ is the optical depth of CSM.
\par
c. $t_{\text{erupt}}$ is the time of the progenitor star erupting
the CSM
shells before explosion. Here we assume the velocity of the progenitor wind $%
v_{\text{w}}=1000$ km s$^{-1}$ and the shells expansion velocities $v_{\text{%
shell}}=100$ km s$^{-1}$.
\end{table*}

\section{Discussion and Conclusions}

\label{sec:dis}

The LC of iPTF15esb has two peaks and a post-peak plateau. All previous
energy-source models (the $^{56}$Ni model, the magnetar model, the
ejecta-CSM model, etc.) cannot account for these exotic features. We suggest
that the LC undulations in iPTF15esb arose from SN ejecta interacting with
dense CSM shells, which may be produced by the eruptions of the progenitor.
Therefore, we generalize the model involving the ejecta and single CSM shell
or wind model to the multiple interaction model. We employ this new model to
fit the LC as well as temperature evolution of iPTF15esb and find it can
well explain the LC. These results indicate that our model is valid.

Based on the modeling, we infer that there are at least three collisions
between the SN ejecta and CSM winds and/or shells and that the progenitor of
iPTF15esb may undergo three discrete mass ejections (two shells and a wind)
before the SN explosion. The pulsational PISN (PPISN) models %
\citep{Woo2007,Woo2017} can account for these eruptions. In these three
successive collisions, the masses of the CSM decrease from the outermost
shell (the first eruption) to innermost wind (the final eruption) but their
density increases. The variation trend of the inferred densities of the
shells and wind is consistent with the stellar structure since the density
of the interior of the star is larger than that of the exterior.

The interaction model for the LC of iPTF15esb is also favored by the broad H$%
\alpha $ emission lines in the late-time spectra which might be produced by
the interaction of SN ejecta with hydrogen-rich CSM shell located at a large
distance from the progenitor star and was ejected by the progenitor star
about 15.5 years before explosion. In our modeling, the first peak of the LC
is powered by the interaction between the SN ejecta and stellar wind while
both the second peak and the plateau are powered by the two CSM shells at
different radii.

The properties of the stellar wind have some implications for the
progenitor. Red supergiants have slow wind velocities of $v_{\text{w,RSG}%
}\approx 10-40$ km s$^{-1}$, with the typical mass loss rate $\dot{M}_{\text{%
RSG}}\approx 10^{-6}-10^{-4}\text{M}_{\odot }$ yr$^{-1}$, while compact
progenitors (e.g., Wolf-Rayet stars) have high wind velocities$\ v_{\text{%
w,WR}}\approx 1000$ km s$^{-1}$, with $\dot{M}_{\text{WR}}\approx
10^{-3}-10^{-1}\text{M}_{\odot }$ yr$^{-1}$ \citep{Smi2014}. By assuming
that the wind velocity $v_{\text{w}}=v_{\text{w,RSG}}$, we can obtain mass
loss rate $\dot{M}=4\pi v_{\text{w}}q_{1}\approx (0.4-1.7)\times 10^{-2}%
\text{M}_{\odot }$ yr$^{-1}$. If $v_{\text{w}}=v_{\text{w,WR}}$, then $\dot{M%
}\approx 0.43\text{M}_{\odot }$ yr$^{-1}$. These results indicate the
progenitor wind of iPTF15esb with an extremely large mass loss rate.

Massive stars are unstable and can lose a lot of matter in the form of
eruptions in the final stage of their lives. Mass loss of the progenitor of
an SN is an important process of stellar evolution. However, our
understanding of the driving mechanism of mass loss is still incomplete.
Further investigations for SLSNe like iPTF15esb should shed light on the
nature of mass-loss history of their progenitors.

\acknowledgments We thank Xue-Feng Wu, Can-Min Deng, and Wei-Kang Zheng for
useful discussions. This work was supported by the National Basic Research
Program (\textquotedblleft 973\textquotedblright\ Program) of China (grant
no. 2014CB845800) and the National Natural Science Foundation of China
(grants nos. 11573014, U1331202, 11533033, and 11673006).

\end{document}